\documentclass[a4paper,aps,prd,10pt,preprintnumbers,showpacs,twocolumn,superscriptaddress,nofootinbib,amsmath,amssymb]{revtex4-1}
\usepackage{graphicx}
\usepackage[utf8]{inputenc}
\usepackage[T1]{fontenc}
\usepackage{cmap}

\def\imo{i}

\def\K{{\cal K}}

\begin{document}
\title{Arbitrarily long-lived quasinormal modes in a wormhole background}
\author{M. S. Churilova}\email{wwrttye@gmail.com}
\affiliation{Institute of Physics and Research Centre of Theoretical Physics and Astrophysics, Faculty of Philosophy and Science, Silesian University in Opava, CZ-746 01 Opava, Czech Republic}
\author{R. A. Konoplya}\email{roman.konoplya@gmail.com}
\affiliation{Institute of Physics and Research Centre of Theoretical Physics and Astrophysics, Faculty of Philosophy and Science, Silesian University in Opava, CZ-746 01 Opava, Czech Republic}
\affiliation{Peoples Friendship University of Russia (RUDN University), 6 Miklukho-Maklaya Street, Moscow 117198, Russian Federation}
\author{A. Zhidenko}\email{olexandr.zhydenko@ufabc.edu.br}
\affiliation{Centro de Matemática, Computação e Cognição (CMCC), Universidade Federal do ABC (UFABC),\\ Rua Abolição, CEP: 09210-180, Santo André, SP, Brazil}
\begin{abstract}
Arbitrarily long lived modes, called quasi-resonances, are known to exist in the spectrum of massive fields for a number of black-hole backgrounds at some discrete values of  mass of the field. Here we show that these modes also exist in the background of wormholes, unless a wormhole has a constant red-shift function, that is, tideless in the radial direction. The evidence of quasi-resonances is supported by calculations in the frequency and time domains, which are in a good concordance. At large masses of the field, time-domain profiles of the absolute value of the wave function have peculiar behavior: the long-lived modes dominate in the signal \textit{after} a long period of power-law tails.
\end{abstract}
\maketitle

\section{Introduction}\label{introduction}

Modern observations of black holes in the electromagnetic and gravitational spectra \cite{Abbott:2016blz,TheLIGOScientific:2016src,Goddi:2017pfy,Akiyama:2019cqa} leave an ample room for interpretations of the geometries of ultracompact objects either in favor of black holes in modified theories of gravity or even towards more exotic objects, such as, for example, stable Schwarzschild stars or wormholes \cite{Damour:2007ap,Konoplya:2016pmh,Konoplya:2019nzp,Camilo:2018goy,Wei:2018aft,Berti:2018vdi,Cardoso:2016rao}. Theoretically, wormholes could be designed in such a way that they would be mimickers of black holes \cite{Damour:2007ap} in various astrophysical optical phenomena and radiation of gravitational waves. At late times classical radiation in the vicinity of compact objects is dominated by the so called quasinormal modes which were extensively studied for black holes and stars \cite{Konoplya:2011qq,Kokkotas:1999bd,Berti:2009kk} and were recently observed in, apparently, the merger of two black holes \cite{Abbott:2016blz,TheLIGOScientific:2016src}. Quasinormal modes of black holes can be described as proper oscillations frequencies of black holes under the following boundary conditions: requirement of purely outgoing wave at infinity and purely incoming wave to the event horizon. In terms of the tortoise coordinate this means absence of incoming waves from both plus- and minus- infinity. The boundary conditions for a traversable wormhole which connects two infinities are the same in terms of the tortoise coordinate \cite{Konoplya:2005et}, so that many of the tools used for finding quasinormal modes of black holes can, with some modifications, be used for wormholes as well \cite{Konoplya:2016hmd,Konoplya:2010kv}. The natural question is then: which of the general features of quasinormal spectrum and scattering properties of black holes are inherited by wormholes? Some of the scattering properties definitely depend more on a particular model of the wormhole or black hole than on the choice between these two objects. However, some features are shared by wormholes in general. For example, in \cite{Konoplya:2016hmd,Konoplya:2010kv} it was observed that amplification of incoming wave owing to rotation, called super-radiance, is possible only for non-symmetric (with respect to the throat) wormholes and only if the asymptotic values of the rotation parameters are different on both sides from the throat. Then, in \cite{Konoplya:2016hmd} it was shown that the symmetric wormholes cannot mimic effectively the ringing of a black hole at a few various dominant multipoles at the same time, so that the future observations of various events should easily tell the symmetric wormhole from the black hole.

An interesting property of black-hole quasinormal spectra takes place for massive fields: when the field mass $\mu$ is increased, the damping rate of the fundamental mode approaches zero at some value of $\mu$, which leads to disappearing of this mode from the spectrum. When mass is further increased, the first overtone behaves in the same way. This phenomenon was first observed in \cite{Ohashi:2004wr} for the massive scalar field in the Reissner-Nordström black-hole background and were called quasi-resonances. In \cite{Konoplya:2004wg} it was analytically shown that the quasi-resonances exist only if the effective potential of the corresponding master wave equation is not zero at infinity. Thus, for instance, the Schwarzschild-de Sitter metric does not allow for such arbitrarily long-lived modes. Further in \cite{Konoplya:2005hr} the existence of quasi-resonances was demonstrated for the massive vector (Proca) field in the Schwarzschild background and for the massive scalar and Dirac fields in Kerr spacetime \cite{Konoplya:2006br,Konoplya:2017tvu}. Quasi-resonances were also found in the spectra of massive fields in higher dimensional spacetimes \cite{Zhidenko:2006rs} and in theories with higher curvature corrections to the Einstein action \cite{Zinhailo:2018ska,Konoplya:2019hml,Zinhailo:2019rwd,Churilova:2019sah}.

Here we would like to learn whether the arbitrarily long-lived quasinormal modes are allowed in the spacetimes of traversable wormholes. In order to answer this question we will study quasinormal modes of a massive scalar field in the background of spherically symmetric Lorenzian traversable wormholes with the help of WKB method and time-domain integration. We shall show that both approaches are in a very good concordance in the common range of applicability and conclude that the wormholes with constant redshift function do not allow for quasi-resonances, while when the redshift function depends on the radial coordinate, there is clear evidence of arbitrarily long-lived modes in the spectrum. In addition, we have found that for sufficiently large masses of the scalar field the long-lived modes dominate not before, but after the period of power-law tails.

The paper is organized as follows. In Sec.~\ref{sec:method} we briefly discuss the WKB method and time-domain integration and show that they can be used for checking existence of quasi-resonances. For this we study the two known cases of a massive scalar field in the Schwarzschild and Schwarzschild-de Sitter backgrounds. For the first case, quasi-resonances appear at some values of mass of the field, while for the second they do not. We will demonstrate that both types of behavior can be clearly seen in the frequency and time domains using the above methods of calculations. In Sec.~\ref{sec:spectra} we analyze spectra of tideless (in the radial direction) wormholes and wormholes with a non-zero tidal force and show that only the second class of wormholes allows for quasi-resonances of the scalar field. In the Conclusion we summarize the obtained results and mention some open questions.

\section{The methods}\label{sec:method}

\subsection{The wave equation}
The metric of a spherically symmetric static spacetime is given by the following general form
\begin{eqnarray}\label{metric}
ds^2 &=& -A(r) dt^2+ B(r){dr^2}+r^2 (\sin^2 \theta d\phi^2+d\theta^2).
\end{eqnarray}
The general covariant equation for a massive scalar field in curved spacetime has the following form
\begin{equation}\label{gensc}
\dfrac{1}{\sqrt{-g}} \partial_\mu(\sqrt{-g} g^{\mu\nu} \partial_\nu\Phi)-\mu^2 \Phi = 0,
\end{equation}
where $\mu$ is mass of the field.
In order to provide the separation of variables, the function $\Phi$ for the scalar field is expressed in terms of the spherical harmonics,
\begin{equation}\label{psi}
\Phi(t,r,y,\phi) = e^{-\imo\omega t}Y_{\ell}(\theta,\phi)\Psi(r)/r,
\end{equation}
where $Y_{\ell}(\theta,\phi)$ are spherical harmonics \cite{harmonics} and $\ell=0,1,2,3\ldots$ is the multipole number.

After separation of angular variables, the wave equation can be represented in the following general form (see, for instance, \cite{Konoplya:2006rv,Zinhailo:2018ska} and references therein):
\begin{equation}  \label{klein-Gordon}
\dfrac{d^2 \Psi}{dr_*^2}+(\omega^2-V(r))\Psi=0,
\end{equation}
where the relation
$$dr_*=dr\sqrt{\frac{B(r)}{A(r)}}$$
defines the ``tortoise coordinate'' $r_*$ and the effective potential is (see e.~g.~Eq.~(12a) in \cite{Zinhailo:2018ska})
\begin{eqnarray}\label{empotential}
V(r) = A(r)\left(\frac{\ell(\ell+1)}{r^2} + \mu^2\right)+\frac{1}{2r}\frac{d}{dr}\frac{A(r)}{B(r)}.
\end{eqnarray}

\subsection{The WKB approach}

For  analysis in the frequency domain we shall use the semi-analytical WKB method \cite{Schutz:1985zz,Iyer:1986np,Matyjasek:2017psv,Konoplya:2003ii,Konoplya:2019hlu}. The essence of this approach is the expansion of the solution at both infinities in the WKB-series and matching of these asymptotic expansions with the Taylor expansion near the peak of the effective potential. In addition, according to \cite{Matyjasek:2017psv}, we use further representation of WKB expansion in the form of the Padé approximants which, in most cases, greatly improves the accuracy of WKB method.

The higher-order WKB formula reads \cite{Konoplya:2019hlu}
\begin{eqnarray}\label{WKBformula-spherical}
\omega^2&=&V_0+A_2(\K^2)+A_4(\K^2)+A_6(\K^2)+\ldots\\\nonumber&-&\imo \K\sqrt{-2V_2}\left(1+A_3(\K^2)+A_5(\K^2)+A_7(\K^2)\ldots\right),
\end{eqnarray}
where $\K=n+1/2$, $n=0,1,2,3\ldots$.

The corrections $A_k(\K^2)$ of order $k$ to the eikonal formula are polynomials of $\K^2$ with rational coefficients and depend on the values $V_2, V_3\ldots$ of higher derivatives of the potential $V(r)$ in its maximum. In order to increase accuracy of the WKB formula, we use the procedure suggested by Matyjasek and Opala \cite{Matyjasek:2017psv}, which consists in usage of the Padé approximants. For the order $k$ of the WKB formula (\ref{WKBformula-spherical}) we define a polynomial $P_k(\epsilon)$ in the following way
\begin{eqnarray}\nonumber
  P_k(\epsilon)&=&V_0+A_2(\K^2)\epsilon^2+A_4(\K^2)\epsilon^4+A_6(\K^2)\epsilon^6+\ldots\\&-&\imo \K\sqrt{-2V_2}\left(\epsilon+A_3(\K^2)\epsilon^3+A_5(\K^2)\epsilon^5\ldots\right),\label{WKBpoly}
\end{eqnarray}
and the squared frequency is obtained for $\epsilon=1$:
$$\omega^2=P_k(1).$$

For the polynomial $P_k(\epsilon)$ we will use Padé approximants
\begin{equation}\label{WKBPade}
P_{\tilde{n}/\tilde{m}}(\epsilon)=\frac{Q_0+Q_1\epsilon+\ldots+Q_{\tilde{n}}\epsilon^{\tilde{n}}}{R_0+R_1\epsilon+\ldots+R_{\tilde{m}}\epsilon^{\tilde{m}}},
\end{equation}
with $\tilde{n}+\tilde{m}=k$, such that, near $\epsilon=0$,
$$P_{\tilde{n}/\tilde{m}}(\epsilon)-P_k(\epsilon)={\cal O}\left(\epsilon^{k+1}\right).$$

Usually for finding fundamental mode ($n=0$) Padé approximants with $\tilde{n}\approx\tilde{m}$ provide the best approximation. In \cite{Matyjasek:2017psv}, $P_{6/6}(1)$ and $P_{6/7}(1)$ were compared to the 6th-order WKB formula $P_{6/0}(1)$. In \cite{Konoplya:2019hlu} it has been observed that usually even $P_{3/3}(1)$, i.~e. a Padé approximation of the 6th-order, gives a more accurate value for the squared frequency than $P_{6/0}(1)$. Here we will use the 4th and 11th WKB expansions with appropriate Padé approximation and show that the results obtained at different WKB orders are in a very good agreement.

\subsection{The time-domain integration}
If one does not use the stationary anzats $\Phi \sim e^{- \imo \omega t}$, but keeps the second derivative in time instead of the $\omega^2$-term, then the integration of the wave equation in the time domain can be used. 
We shall integrate the wave-like equation rewritten in terms of the light-cone variables $u=t-r_*$ and $v=t+r_*$. The appropriate discretization scheme was suggested in \cite{Gundlach:1993tp}:
\begin{eqnarray}\label{Discretization}
\Psi\left(N\right)&=&\Psi\left(W\right)+\Psi\left(E\right)-\Psi\left(S\right)-\\\nonumber
&&-\Delta^2\frac{V\left(W\right)\Psi\left(W\right)+V\left(E\right)\Psi\left(E\right)}{8}+{\cal O}\left(\Delta^4\right)\,,
\end{eqnarray}
where we used the following designations for the points:
$N=\left(u+\Delta,v+\Delta\right)$, $W=\left(u+\Delta,v\right)$, $E=\left(u,v+\Delta\right)$ and $S=\left(u,v\right)$. The initial data are given on the null surfaces $u=u_0$ and $v=v_0$.  In order to extract the values of the quasinormal frequencies we will use the Prony method which allows us to fit the signal by a sum of exponents with some excitation factors.

\subsection{Test of applicability of the methods to calculation of long-lived modes}

In order to understand whether the chosen methods are adequate for quasinormal modes with very small imaginary part we will first test them for the two cases which were previously studied with the accurate, convergent (Frobenius) method in \cite{Ohashi:2004wr,Konoplya:2004wg}. The first case is a massive scalar field in the Schwarzschild spacetime which allows for quasi-resonances and the second case is asymptotically de Sitter Schwazrschild black hole for which the quasi-resonances are forbidden \cite{Konoplya:2004wg}.

The metric functions for these cases are given in the following form:
\begin{equation}
A(r) = \frac{1}{B(r)} = 1 - \frac{2 M}{r} - \frac{\Lambda r^2}{3},
\end{equation}
where $M$ is the black-hole mass and $\Lambda$ is the cosmological constant.

\begin{table*}
\begin{tabular}{p{1cm}cccc}
\hline
$\mu$ & $\Lambda=0$ WKB & $\Lambda=0$ (time domain) & $\Lambda=0.1$ WKB & \hspace{.2cm} $\Lambda=0.1$ (time domain) \\
\hline
$0   $ & $4.04264 - 0.192512\imo~$ & $4.043 - 0.192\imo$ & $3.55927 - 0.169453\imo$ & $3.559459 - 0.169425\imo$ \\
$1   $ & $4.08373 - 0.189241\imo~$ & $4.084 - 0.189\imo$ & $3.59550 - 0.167232\imo$ & $3.595690 - 0.167204\imo$ \\
$2   $ & $4.20788 - 0.179298\imo~$ & $4.208 - 0.179\imo$ & $3.70426 - 0.160709\imo$ & $3.704475 - 0.160681\imo$ \\
$3   $ & $4.41792 - 0.162229\imo~$ & $4.418 - 0.162\imo$ & $3.88574 - 0.150374\imo$ & $3.885986 - 0.150344\imo$ \\
$4   $ & $4.71958 - 0.136962\imo~$ & $4.720 - 0.137\imo$ & $4.13976 - 0.137280\imo$ & $4.140058 - 0.137249\imo$ \\
$5   $ & $5.12409 - 0.100639\imo~$ & $5.125 - 0.101\imo$ & $4.46479 - 0.123345\imo$ & $4.465164 - 0.123316\imo$ \\
$6   $ & $5.65527 - 0.035481\imo$ & $5.663 - 0.034\imo$ & $4.85608 - 0.111228\imo$ & $4.856556 - 0.111201\imo$ \\
$6.01$ & $5.66028 - 0.033719\imo$ & $5.670 - 0.033\imo$ & $4.86030 - 0.111124\imo$ & $4.860777 - 0.111097\imo$ \\
$6.02$ & $5.66471 - 0.031765\imo$ & $5.676 - 0.032\imo$ & $4.86452 - 0.111021\imo$ & $4.865003 - 0.110994\imo$ \\
$6.03$ & $5.66813 - 0.029561\imo$ & $5.683 - 0.031\imo$ & $4.86875 - 0.110918\imo$ & $4.869235 - 0.110891\imo$ \\
\hline
\end{tabular}
\caption{Fundamental ($n=0$) quasinormal modes ($\ell=10$) for Schwarzschild ($\Lambda=0$) and Schwarzschild-de Sitter ($\Lambda=0.1$) black holes obtained by the 4th order WKB ($\tilde{n}=3$, $\tilde{m}=1$) and time-domain methods in the units of half mass ($M=1/2$).}\label{tabl:SdS}
\end{table*}

As can be seen from Table~\ref{tabl:SdS}, for sufficiently large multipole numbers $\ell$ both methods are in a very good agreement and both indicate that the imaginary part of the fundamental quasinormal mode goes to zero in the Schwarzschild background and to some non-zero constant in the Schwarzschild-de Sitter case.

\section{Quasi-resonances in the background of wormholes}\label{sec:spectra}

Here we will start from rather an agnostic point of view and consider some wormhole metrics which represent Lorenzian traversable wormholes, and will not investigate the fundamental issues of viability of these models related to the possible underlying gravitational theory, equation of state of matter supporting the wormhole or its stability against gravitational perturbations. We justify such an approach, because we are interested in some basic feature of a massive field (quasi-resonance) propagating in a wormhole's geometry and not interested in a particular dependence of this characteristic on the design of the wormhole and the matter supporting it.

The spacetime of a generic  static spherically symmetric wormhole can be described with the help of the following metric functions \cite{Morris-Thorne}:
\begin{equation}
A(r) = e^{2\Phi(r)}, \quad B^{-1}(r) = {1-\dfrac{b(r)}{r}},
\end{equation}
where $\Phi(r)$ is called the redshift function and $b(r)$ is the shape function. When $\Phi(r) = 0$, there is no tidal force in the radial direction.
As a simple example of a wormhole we will consider the following case:
\begin{equation}
e^{2\Phi(r)}=1- \frac{2M}{r}, \qquad b(r)=\frac{q^2}{r},
\end{equation}
which goes over into the Ellis-Bronnikov wormhole \cite{Bronnikov-Ellis} in the limit $M\to0$.
Here $r^{*}$ is the tortoise coordinate with two branches (positive and negative) defined as
\begin{equation}
r^{*}(r)=\pm\intop_q^r\frac{dr}{e^{\Phi(r)}\sqrt{1-\frac{b(r)}{r}}},
\end{equation}
and the effective potential is (\ref{empotential})
$$V(r)=e^{2\Phi(r)}\left(\frac{\ell(\ell+1)}{r^2}+\mu^2\right)+\frac{1}{2r}\frac{d}{dr}e^{2\Phi(r)}\left(1-\frac{b(r)}{r}\right).$$
Thus the effective potential is symmetric with respect to the wormhole's throat, which is situated at $b(r)=r$.
One can easily see that when $\Phi(r) = 0$ the spectrum of the massless scalar field is acquiring the shift
\begin{equation}
\omega^2 \rightarrow \omega^2 - \mu^2,
\end{equation}
so that the modes with zero imaginary part is achieved only in the limit $\mu \rightarrow \infty$. In other words, the quasi-resonances do not appear in the spectrum of the tideless wormhole.

\begin{table*}
\begin{tabular}{p{2cm}cc}
\hline
$\mu$ & 4th order WKB ($\tilde{n}=3$, $\tilde{m}=1$) &\hspace{.2cm} 11th order WKB ($\tilde{n}=6$, $\tilde{m}=5$)\\
\hline
$ 0   $ & $9.39452 - 0.295939\imo$ & $9.39452 - 0.295938\imo$ \\
$ 4   $ & $10.0520 - 0.273704\imo$ & $10.0520 - 0.273703\imo$ \\
$ 8   $ & $11.8075 - 0.225506\imo$ & $11.8075 - 0.225504\imo$ \\
$12   $ & $14.2615 - 0.175858\imo$ & $14.2615 - 0.175856\imo$ \\
$16   $ & $17.1162 - 0.132854\imo$ & $17.1162 - 0.132852\imo$ \\
$20   $ & $20.2023 - 0.095647\imo$ & $20.2023 - 0.095645\imo$ \\
$24   $ & $23.4275 - 0.060232\imo$ & $23.4275 - 0.060231\imo$ \\
$25   $ & $24.2487 - 0.050826\imo$ & $24.2487 - 0.050816\imo$ \\
$26   $ & $25.0743 - 0.040742\imo$ & $25.0742 - 0.040795\imo$ \\
$26.25$ & $25.2813 - 0.038002\imo$ & $25.2811 - 0.038163\imo$ \\
$26.5 $ & $25.4885 - 0.035064\imo$ & $25.4882 - 0.035484\imo$ \\
$26.75$ & $25.6957 - 0.031804\imo$ & $25.6954 - 0.032752\imo$ \\
$27   $ & $25.9027 - 0.028041\imo$ & $25.9027 - 0.029926\imo$ \\
$27.25$ & $26.1084 - 0.023532\imo$ & $26.1096 - 0.027379\imo$ \\
$27.5 $ & $26.3096 - 0.017819\imo$ & $26.3172 - 0.028294\imo$ \\
$27.75$ & $26.4663 - 0.009098\imo$ & $26.5397 - 0.020447\imo$ \\
$28   $ & $26.7979 - 0.000000\imo $ & $26.7710-0.000000\imo $  \\
\hline
\end{tabular}
\caption{Fundamental ($n=0$) quasinormal modes for the wormhole ($M=0.1$, $q=1$) obtained by the WKB method, $\ell=10$.}\label{tabl:wormhole:l10}
\end{table*}

\begin{table*}
\begin{tabular}{p{1cm}ccc}
\hline
$\mu$ & 4th order WKB ($\tilde{n}=3$, $\tilde{m}=1$) &\hspace{.2cm} 11th order WKB ($\tilde{n}=6$, $\tilde{m}=5$) & time domain\\
\hline
$0  $ & $1.35968 - 0.303036\imo$ & $1.36168 - 0.301228\imo$ & $1.36224 - 0.302098\imo$\\
$0.2$ & $1.37073 - 0.300245\imo$ & $1.37275 - 0.298449\imo$ & $1.36325 - 0.296535\imo$\\
$0.4$ & $1.40346 - 0.292220\imo$ & $1.40546 - 0.290457\imo$ & $1.39818 - 0.292720\imo$\\
$0.6$ & $1.45667 - 0.279896\imo$ & $1.45866 - 0.278178\imo$ & $1.45176 - 0.279843\imo$\\
$0.8$ & $1.52854 - 0.264524\imo$ & $1.53052 - 0.262851\imo$ & $1.52376 - 0.262224\imo$\\
$1  $ & $1.61685 - 0.247366\imo$ & $1.61883 - 0.245748\imo$ & \\
$1.2$ & $1.71924 - 0.229488\imo$ & $1.72123 - 0.227906\imo$ & \\
$1.4$ & $1.83346 - 0.211664\imo$ & $1.83545 - 0.210099\imo$ & \\
$1.6$ & $1.95745 - 0.194392\imo$ & $1.95946 - 0.192821\imo$ & \\
$1.8$ & $2.08947 - 0.177945\imo$ & $2.09151 - 0.176343\imo$ & \\
$2  $ & $2.22803 - 0.162441\imo$ & $2.23012 - 0.160769\imo$ & \\
$2.2$ & $2.37193 - 0.147899\imo$ & $2.37411 - 0.146122\imo$ & \\
$2.4$ & $2.52018 - 0.134290\imo$ & $2.52244 - 0.132343\imo$ & \\
$2.6$ & $2.67198 - 0.121561\imo$ & $2.67433 - 0.119356\imo$ & \\
$2.8$ & $2.82666 - 0.109671\imo$ & $2.82914 - 0.107078\imo$ & \\
$3  $ & $2.98370 - 0.098599\imo$ & $2.98661 - 0.095540\imo$ & \\
$3.2$ & $3.14276 - 0.088347\imo$ & $3.14566 - 0.084946\imo$ & \\
$3.4$ & $3.30376 - 0.078811\imo$ & $3.30602 - 0.075141\imo$ & \\
$3.6$ & $3.46593 - 0.070935\imo$ & $3.46655 - 0.068118\imo$ & \\
$3.8$ & $3.62958 - 0.060684\imo$ & $3.62827 - 0.060145\imo$ & \\
$4  $ & $3.79117 - 0.048620\imo$ & $3.79013 - 0.059423\imo$ & \\
\hline
\end{tabular}
\caption{Fundamental ($n=0$) quasinormal modes for the wormhole ($M=0.1$, $q=1$) obtained by the WKB method, $\ell=1$.}\label{tabl:wormhole:l1}
\end{table*}

In the case of a wormhole with non-constant redshift function $\Phi(r)$ from Tables~\ref{tabl:wormhole:l10}~and~\ref{tabl:wormhole:l1}  one can see that the real oscillation frequency monotonically increases and the damping rate decreases when the mass of the field $\mu$ is growing. The data  with the help of the 4th and 11th WKB orders are in a very good agreement which means that the  results do not depend on the order of WKB formula qualitatively. Although the extrapolation of the WKB data alone indicates the existence of quasi-resonances, we tried to explore the time-domain integration in this case as well and found a peculiar behavior. For small and moderate values of $\mu$ the period of quasinormal ringing occurs relatively early in the time-domain profile. While for sufficiently large $\mu$ the power-law tail appears first in the profile and later is changed by the long-lived quasinormal mode. Therefore, the regime of long-lived quasinormal modes is difficult to study via the time-domain integration: one needs superior accuracy, very small grid of integration and approaching very late times at which the long-lived modes become dominating.

\begin{figure*}
\resizebox{\linewidth}{!}{\includegraphics*{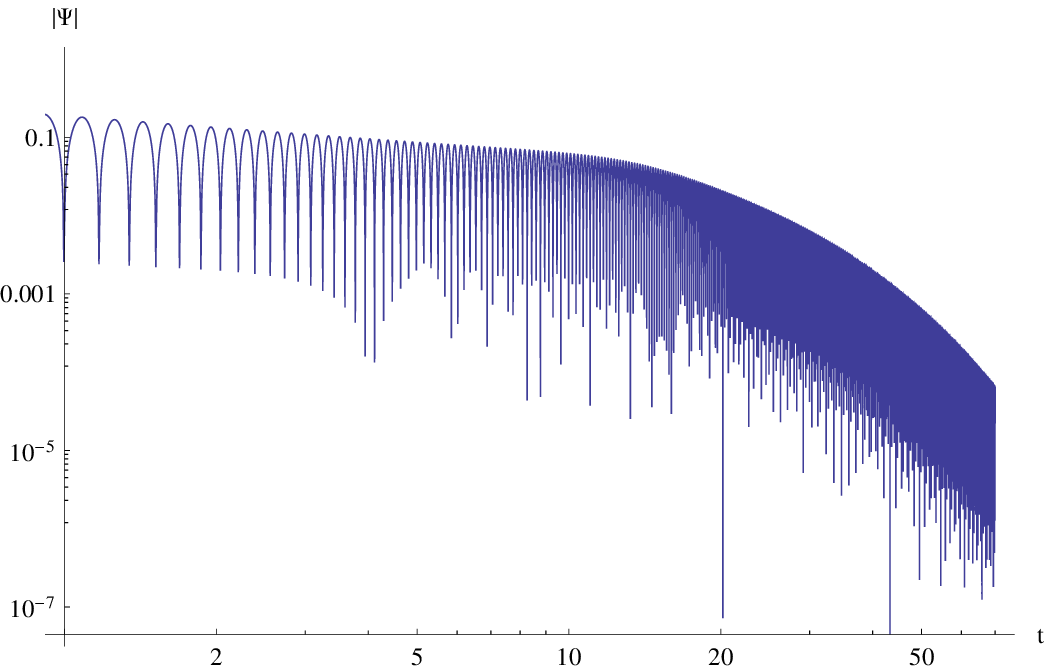}\includegraphics*{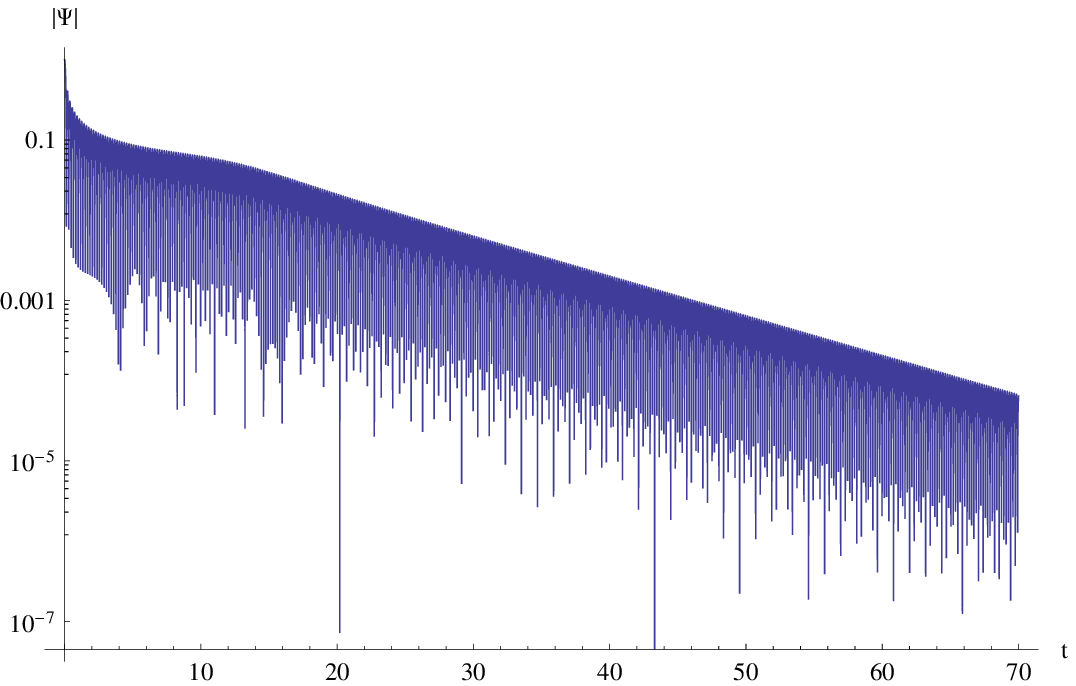}}
\caption{Time-domain profile in the logarithmic (left) and semi-logarithmic (right) scales for a massive scalar field ($\mu =18$, $\ell=10$) in the wormhole spacetime ($M=0.1$, $q=1$). The initial power-law tail is followed by the exponentially decaying long-lived quasinormal mode.}\label{fig1}
\end{figure*}

In fig.~\ref{fig1} we can see that at late times the power-law tail goes over into the exponential decay dominated by the long-lived quasinormal modes. The WKB data for this mode $\omega_{WKB} = 18.6374 - 0.113718 i$ is very close to what we observe in the time-domain $\omega_{TD} = 18.6381 - 0.1137 i$. However, in order to continue these calculations to really small values of the damping rate in the time domain, enormous computer time is necessary. The full understanding of this non-trivial behavior of quasi-resonances in time-domain, for both wormholes and, if it occurs, also for black holes, must be analyzed in a more rigorous approach by consideration the excitation factors and Green functions of the corresponding wave equation. We believe that existence of quasi-resonances is a general property of asymptotically flat wormholes with non-zero radial tidal force and many more examples of this could be collected.

\section{Conclusion}

Quasinormal modes of wormholes have been recently studied in a number of papers in the context of possible astrophysical observations, testing wormhole geometry and stability
\cite{Roy:2019yrr,Ovgun:2019yor,Bronnikov:2012ch,Kim:2018ang,Oliveira:2018oha,Blazquez-Salcedo:2018ipc,Cuyubamba:2018jdl,Konoplya:2018ala,Aneesh:2018hlp,Bronnikov:2019sbx}. Here we were interested in the question whether the arbitrarily long-lived quasinormal modes, which exist in the spectra of massive fields in some black-hole backgrounds, are also present in the spectra of wormholes? We have shown that wormholes with a constant redshift function do not allow for such arbitrarily long-lived modes, while wormholes with non-vanishing radial tidal force do allow for quasi-resonances.
We have observed interesting behavior of the time-domain profile of a massive scalar field in the regime of large mass $\mu$: the power-law tail which usually appears at asymptotically late times, dominates in the signal before the exponentially decaying long-lived mode, so that, in a sense, the roles of power-law tails and quasinormal modes are interchanged. We believe that mathematically strict description of this regime must be possible when analyzing behavior of excitation factors and Green functions of the perturbation.

In addition to fundamental massive fields, a number of scenarios may lead to appearing of an effective massive mode: for example, due to additional spacial dimensions in brane-world models or owing to a non-minimal coupling of the matter field to gravity. Then, the fact that arbitrarily long lived modes can appear at some values of the effective mass might be important for observations. The appearance of the power-law tails \textit{before} the long-lived quasinormal modes could be an indicator allowing one to distinguish a wormhole from a black hole.

\acknowledgments{
The authors acknowledge the support of the grant 19-03950S of Czech Science Foundation (GAČR). A.~Z. was supported by Conselho Nacional de Desenvolvimento Científico e Tecnológico (CNPq).
}

\end{document}